\documentclass[11pt, a4]{article}

\usepackage{amsmath}
\usepackage{caption}
\usepackage{hyperref}
\usepackage{bm}
\usepackage{authblk}
\usepackage[psamsfonts]{amsfonts,amssymb}
\usepackage{amsthm,enumerate}
\usepackage{color}
\usepackage{tikz} 
\usepackage{pgfplots}
\usepackage{graphicx}
\usepackage[none]{hyphenat}

\title{{\sf Epilocal}: a real-time tool for local epidemic monitoring}

\author[1]{Marco Bonetti\thanks{Corresponding author: \url{marco.bonetti@unibocconi.it}\\ We wish to thank Marcello Pagano, Elena Savoia, Rino Bellocco and Alessia Melegaro for useful discussions and comments on this work.}}
\author[2,3]{Ugofilippo Basellini}

\affil[1]{Carlo F. Dondena Research Center and Bocconi Institute for Data Science and Analytics, Bocconi University, Milan, Italy}
\affil[2]{Max Planck Institute for Demographic Research (MPIDR), Rostock, Germany}
\affil[3]{Institut national d'\'{e}tudes d\'emographiques (INED), Aubervilliers, France}

\date{DRAFT April 16, 2020}

\begin{document}
\captionsetup[figure]{labelfont={bf},name={Figure},labelsep=period}
\maketitle

\abstract{We describe {\sf Epilocal}, a simple \textsf{R} program designed to automatically download the most recent data on reported infected SARS-CoV-2 cases for all Italian provinces and regions, and to provide a simple descriptive analysis. For each province the cumulative number of reported infected cases is available each day. In addition, the current numbers of hospitalized patients (separately for intensive care or not) and the cumulative number of deceased individuals are available at the region level. The data are analyzed through Poisson generalized linear models with logarithmic link function and polynomial regression on time. For cumulative data, we also consider a logistic parameterisation of the hazard function. Automatic model selection is performed to choose among the different model specifications, based on the statistical significance of the corresponding estimated parameters and on goodness-of-fit assessment. The chosen model is used to produce up-to-today estimates of the growth rate of the counts. Results are plotted on a map of the country to allow for a visual assessment of the geographic distribution of the areas with differential prevalence and rates of growth.}

\section{Introduction}
The SARS-CoV-2 virus (also known as coronavirus) has emerged between the end of 2019 and the beginning of 2020 as a global threat. The disease caused by the virus has been designated COVID-19. It is hitting most countries in the world at a very quick rate, and with differential timing both across countries and within countries (see, e.g. \cite{who:2020}). Such differential spreading of the epidemic over different regions provides the opportunity for optimizing the use of the health care resources across a country, thus potentially reducing mortality due to lack of sufficient care.

We have developed a simple software tool to allow for the quick monitoring of province-level and region-level data on the cumulative number of reported infected cases in Italy, as well as other cumulative and non-cumulative counts. In Section 2 we describe the methods, and in particular the data sources, the automatic model selection process, and the output provided by the software. In Section 3 we show a few sample plots for some Italian provinces and regions, as well as the graphical representation for the whole country. We close with some discussion in Section 4.

\section{Methods}
\subsection{Data sources and modeling}
Daily data on the reported numbers of infected individuals with the SARS-CoV-2 virus are available at the province- and region-level in Italy from the Department of Civil Protection \cite{dpc:2020}. The data report the daily cumulative number for each of the 107 provinces and 20 regions starting from February 24th, 2020, together with representative coordinates (longitude and latitude) for each geographical unit. Similar daily data are available, only at the region-level, for the cumulative number of Covid-positive, deceased individuals. The resident population of each province and region on January 1, 2019 was obtained from the Italian Statistical Institute \cite{istat:2020}. The number of positive individuals who are currently present in intensive care unit (ICU) and non-ICU departments are available at the regional level. In the description below we will focus on the cumulative number of reported positive cases for a given province, but the software can produce similar analyses for all other counts at the region-level, with only minor differences which we will describe in our closing section.

Given the resident population $N$ (assumed constant) for a generic province, we let $Y_t$ be the number of reported infected at time $t$ in that province. We model the observed number of counts ($Y_t$) via a Poisson generalized linear model with an hazard rate $h(t)$, the natural logarithm link function and the exposed population ($N$) as an offset \cite{mccu:neld:1989}. For the hazard rate, we consider some parametric functions of time, specifically first-, second-, or third-degree (log) polynomials for all responses, and a logistic parameterisation only for cumulative counts. Time is a natural number between the day when the number of reported case reached 10 and the most recent day for which data are available

Let us focus on the third-degree log-polynomial hazard model, i.e.~$\eta(t)=\log (h(t)) = \beta_0 + \beta_1 t + \beta_2 t^2 + \beta_3 t^3$. We thus assume $Y_t \sim {\rm Poisson}(\lambda_t)$, with
\begin{equation}
\begin{aligned}
   \log(\lambda_t) &= \log(N) + \eta(t) \\
   &= \log(N) + \beta_0 + \beta_1 t + \beta_2 t^2 + \beta_3 t^3. 
\end{aligned}   
\end{equation}
Note the inclusion of the offset term $\log(N)$ to explicitly extract that (non-time-varying) exposure term from the constant parameter of the regression.

The vector parameter ${\bm{\beta}}=(\beta_0, \beta_1, \beta_2, \beta_3)^T$ is estimated by maximum likelihood in \textsf{R}  \cite{R:2013}. We do not fit the model for provinces that have experienced a maximum reported infected rate smaller than $0.00005$, and just output the ''Prevalence too low'' message for those cases.

The predicted number of reported infected cases at time $t$ is thus

\begin{equation}
\widehat{Y}_t = \widehat{\lambda}_t  = N \, h(t)  = N \exp \left[ \widehat{\beta}_0 + \widehat{\beta}_1 t + \widehat{\beta}_2 t^2 +\widehat{\beta}_3 t^3 \right]. \label{lambdahat}
\end{equation}


We can easily compute the first derivative of the rate $h(t)$ and estimate it by plugging-in the maximum likelihood estimator (mle) $\widehat{\bm{\beta}}$ into the expression, and we obtain
\begin{equation}
\widehat{\frac{\partial}{\partial t} \, h(t)} =  \widehat{h}(t)  \left( \widehat{\beta}_1 + 2 \widehat{\beta}_2 t + 3 \widehat{\beta}_3 t^2 \right) .
\end{equation}

Similarly, the second derivative (useful to identify inflection points) is estimated by

\begin{equation}
\widehat{\frac{\partial d^2}{\partial t^2} \, h(t) } = \widehat{h}(t) \left[ \left( 2 \widehat{\beta}_2 + 6 \widehat{\beta}_3 t \right) + \left(  \widehat{\beta}_1 + 2 \widehat{\beta}_2 t + 3 \widehat{\beta}_3 t^2 \right)^2 \right] .  \label{secder3}
\end{equation}
Note that, by invariance, these estimators are the mles of the quantities that they are estimators for. We did not pursue it here, but their large-sample distribution can be obtained by the delta method from the approximate distribution of $\widehat{\bm{\beta}}$. In particular, they are also approximately normally distributed, and a consistent estimator for the asymptotic variance can easily be constructed, if desired.

While we have described the third-degree polynomial model above, the first-degree and the second-degree models follow immediately by dropping either both the quadratic and the cubic term, or just the cubic term from the third-degree model. The predicted rates for these models then follow immediately, and their estimated trends and curvatures can be easily shown to be equal to

\begin{equation}
\widehat{\frac{\partial}{\partial t} \, h(t)} =  \widehat{h}(t) \left( \widehat{\beta}_1 + 2 \widehat{\beta}_2 t \right) ; \ \  \widehat{\frac{\partial d^2}{\partial t^2} \, h(t)} = \widehat{h}(t) \left[ 2 \widehat{\beta}_2  + \left(  \widehat{\beta}_1 + 2 \widehat{\beta}_2 t  \right)^2  \right]
\end{equation}
for the second-degree model, and 

\begin{equation}
\widehat{\frac{\partial}{\partial t} \, h(t)} =  \widehat{h}(t) \, \widehat{\beta}_1; \ \ \widehat{\frac{\partial d^2}{\partial t^2} \, h(t) } =  \widehat{h}(t) \, \widehat{\beta}_1 ^2
\end{equation}
for the first-degree model. Note that the expression of $\widehat{\lambda}_t$ should be adjusted by dropping the appropriate terms from Equation \ref{lambdahat} above. Also, one can trivially obtain the expressions for the corresponding first and second derivatives of $Y_t$ by multiplying the formulas by $N$, i.e.~$\widehat{\frac{\partial}{\partial t} \,Y_t} = N \widehat{\frac{\partial}{\partial t} \,h(t)}$. 

These formulas allow one to estimate the rate of change in the reported infected rate in particular at the most recent time point. Below we will exploit them to produce a map of the provinces that reflects such trends.

Note that the formulas for the first derivatives also allow for the immediate calculation of the time(s) $\tilde{t}$ at which the local epidemic will reach a maximum (or minimum) for the second-degree model and for the third-degree model. These local maxima (or minima) are, for the two models,

\begin{equation}
\tilde{t}= - \frac{\widehat{\beta}_1}{2 \widehat{\beta}_2} \ \ {\rm and} \ \ \tilde{t}_{1,2} = \frac{-2 \widehat{\beta}_2  \pm \left[ 4  \widehat{\beta}_2^2 - 12 \widehat{\beta}_1 \widehat{\beta}_3 \right]^{1/2}}{6 \widehat{\beta}_3},
\end{equation}
when $4  \widehat{\beta}_2^2 - 12 \widehat{\beta}_1 \widehat{\beta}_3 >0$.
These predicted local or global maxima and minima are likely to be quite unstable until enough data are collected, and they should be interpreted with caution. As time goes by, one expects such estimated times to have increasing precision. More generally, due to the polynomial shape that we have chosen for the regression component, we expect these models to only be adequate to describe the epidemic during the initial growth phase for the first-degree model, and only up until the maximum prevalence of the epidemic for the second-degree and third-degree models. In particular, any predicted decreases in the cumulative number of cases is clearly meaningless since recovered cases are not accounted for in the data. In general, any forecasting beyond the last observed time point should be performed with caution. 

Cumulative counts can only increase, and they can be expected to reach a plateau. As a consequence, we have also considered a logistic curve as a way to describe such cumulative counts. The functional form of the hazard function is 
\begin{equation}
    h(t) = K \frac{\exp \left( \beta_0 + \beta_1 t \right)}{1+\exp\left(\beta_0 + \beta_1 t \right)} ,
\end{equation}
where $K = \exp(\gamma) / \left[ 1+ \exp(\gamma) \right] $ is a parameter that takes value in $(0, 1)$, corresponding to the upper asymptote of the logistic curve.  Since the logistic (log) hazard is a non-linear function of the parameters $\bm{\theta} = \left[K, \beta_0, \beta_1 \right]$, we estimate the model's parameters by maximising the Poisson log-likelihood
\begin{equation}
\ln \, \mathcal{L}\left(\bm{\theta}\,|\, Y_{t}\,,N\, \right) \propto \sum_{t} \left[  Y_{t} \,
\ln \left(h(\bm{\theta}; t)\right) - N
\, h(\bm{\theta}; t)  \right]  \, .
\end{equation}
In \textsf{R}, this can be performed using the \textsf{maxLik} package \cite{henningsen2011maxLik}.

As for the log-polynomial models, the slope and the curvature of the logistic model at a given time $t$ can be computed from the first and second derivatives with respect to time
\begin{equation}\label{Eq:LogisSecDer}
\begin{aligned}
    \widehat{\frac{\partial}{\partial t} \, h(t) } &= \widehat{K} \, \widehat{\beta}_1 \frac{\exp\left(\widehat{\beta}_0 + \widehat{\beta}_1 t \right)}{\left[1+ \exp\left( \widehat{\beta}_0 + \widehat{\beta}_1 t\right) \right]^2} \\
    \widehat{\frac{\partial d^2}{\partial t^2} \, h(t) } &= \widehat{K} \, \widehat{\beta}_1^2 \frac{\exp\left(\widehat{\beta}_0 + \widehat{\beta}_1 t \right) \left[1- \exp\left( \widehat{\beta}_0 + \widehat{\beta}_1 t\right) \right]}{\left[1+ \exp\left( \widehat{\beta}_0 + \widehat{\beta}_1 t\right) \right]^3}.
\end{aligned}
\end{equation}

An important feature of the logistic function, which we will exploit for the model selection, is its inflection point. This can be computed by letting the second derivative be equal to zero
\begin{equation}\label{Eq:LogisSecDer}
    \widehat{\frac{\partial d^2}{\partial t^2} \, h(t) } = 0 \quad \Leftrightarrow \quad t = - \frac{\widehat{\beta}_0}{\widehat{\beta}_1}
\end{equation}

Finally, we compute 95\% confidence intervals for the inflection point by delta method, i.e.~from the gradient of $h(t)$ with respect to  its parameters and the variance-covariance matrix of the estimated parameters. 


\subsection{Automatic model selection}
\label{graphics}
The likelihood maximization algorithm might not converge, and that needs to be monitored and accounted for in the model selection process described below.
\newline
Specifically, we initially try to fit the three nested models (first-, second-, and third-degree) to the data, and recover the convergence status for all three models. We then analyze the 8 possible cases separately. If only one of the models converges, we select that model. If two models converge, we consider the significance of the larger model through either the appropriate Wald test statistic (if the two models differ by just one term) or by the likelihood ratio test statistic if the two models that converge are the first-degree and the third-degree models. We then select the best model between the two.
\newline
If all three models converge, we perform a backward elimination process starting from the third-degree polynomial model: if that is significantly better than the second-degree polynomial we select the former; if not, we drop the third-degree model and compare the second-degree model to the first-degree model, and select the final model between those. For all model selection tests we use a p-value cutoff equal to $0.05$.

When dealing with cumulative counts, it is important to consider the logistic model. In general, we expect the polynomial model to provide an accurate fit at the beginning and in the middle of the growth of the epidemic, but we expect the logistic curve to provide a better fit in the latest stages of a pandemic. In particular, the polynomial model tends to mimic a logistic-type curve after the response variable starts to flatten. When the selected polynomial model indicates a downward trend at the current date, we consider this an indication that a logistic pattern has been achieved. However, in order to avoid selecting the logistic curve too early in time, we retain such model only if the 95\% confidence interval for estimated inflection point (flex) falls entirely to the left of 7 days prior to the last observed data point. Should that criterion not be met, we simply report that no adequate model could be found.  


\subsection{Output and graphical displays}
We first produce a separate plot for each province and region. In each plot, we show the reported cases over time together with a line showing the fitted values from the selected model. The plot also contains the indication of the degree of the polynomial that was selected (or the logistic model for cumulative responses), and the estimated rate of increase at the last time point, and the possible messages signaling that no modeling was performed due to the low prevalence in that province, or for lack of convergence of all models.
\newline
We then plot the estimated slopes on a map of Italy as obtained from google maps using the {\sf ggmap} package (\cite{google:2020};\cite{kahl:wick:2013}). For each province or region, we plot an empty circle centered at the representative coordinates, with the area of the circle proportional to the estimated prevalence at the most recent time, i.e.~$\widehat{Y}_t/N$. When no appropriate model is found, we average the last two observed reported positives over the resident population.
\newline
Furthermore, we use color-coded circles to indicate the value of the growth rates, estimated as described in Subsection 2.1. We employ the perceptually uniform color palette from the {\sf viridis} package to map the growth rates to each circle \cite{garnier2018viridis}.
\newline
Lastly, the text output contains the observed rate of cases over the population; all the details about the models that were fit to each province; the estimated maxima / minima; and a final tables containing the names of all provinces and the estimated slope at the last time point, together with the class used to color the circle in the country map.


\section{Results}
We ran our software on the most recent data (the last available data point is indicated on the plots). The complete output contains one plot for each province, one figure for the whole country, and a detailed text output (not shown here).

\begin{figure}[h!]
\centering
\includegraphics[width = 6.5in]{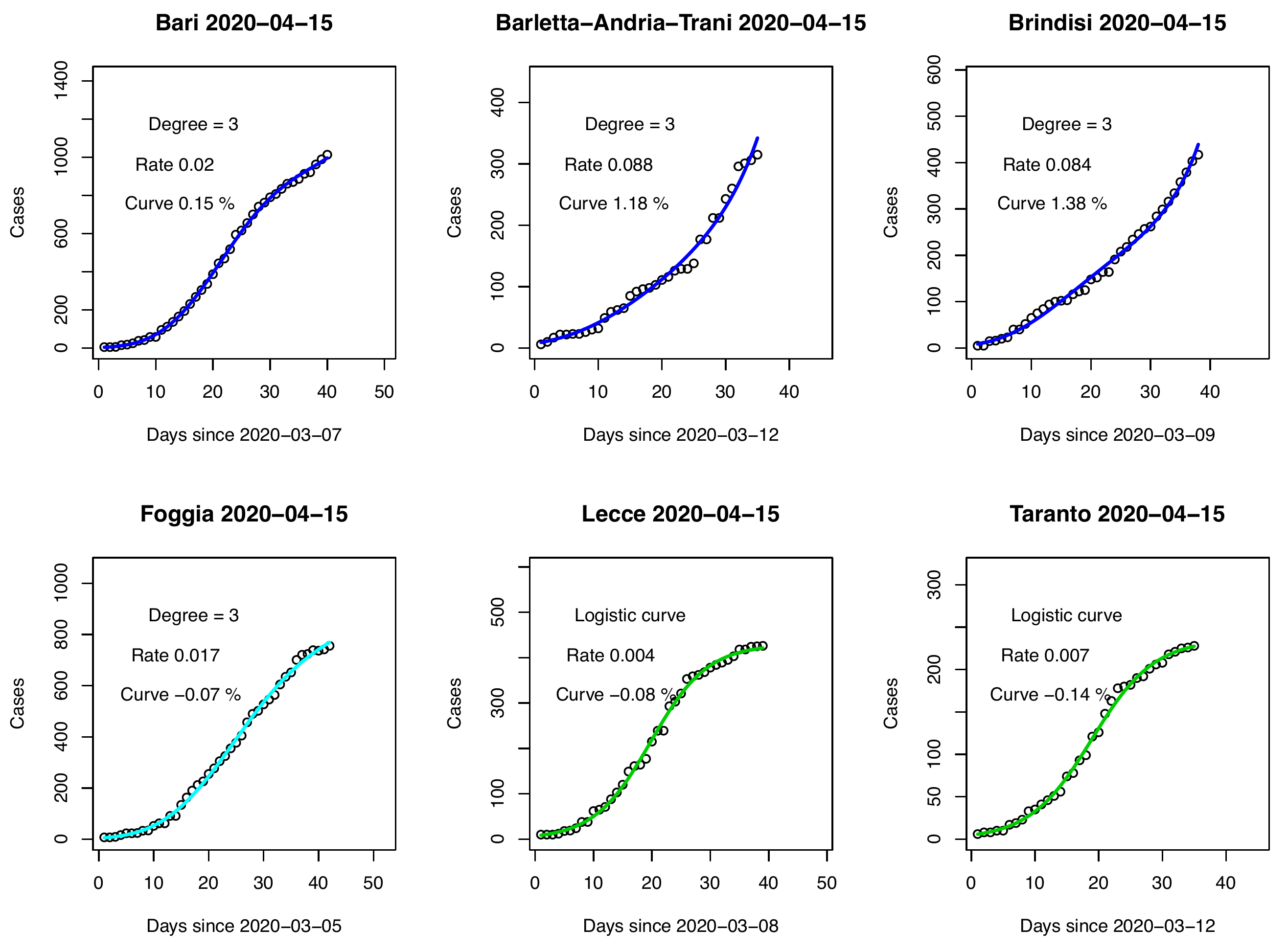}
\caption{{\small Sample of plots for 6 Italian provinces, showing the cumulative number of reported infected SARS-CoV-2 cases over time since the day when the reported cases reached the number of 5. The curves show the fitted values from the model selected for each province.}}
\label{fig:4plots}
\end{figure}

Figure \ref{fig:4plots} shows a representative example of some province-specific plots, with the annotated date of last data point in the title, the degree and kind of selected model, and any warnings. Probably also due to our use of cumulative counts, the observed fit is clearly very good, even just from visual inspection. 

\begin{figure}[h!]
\centering
\includegraphics[width = 6in]{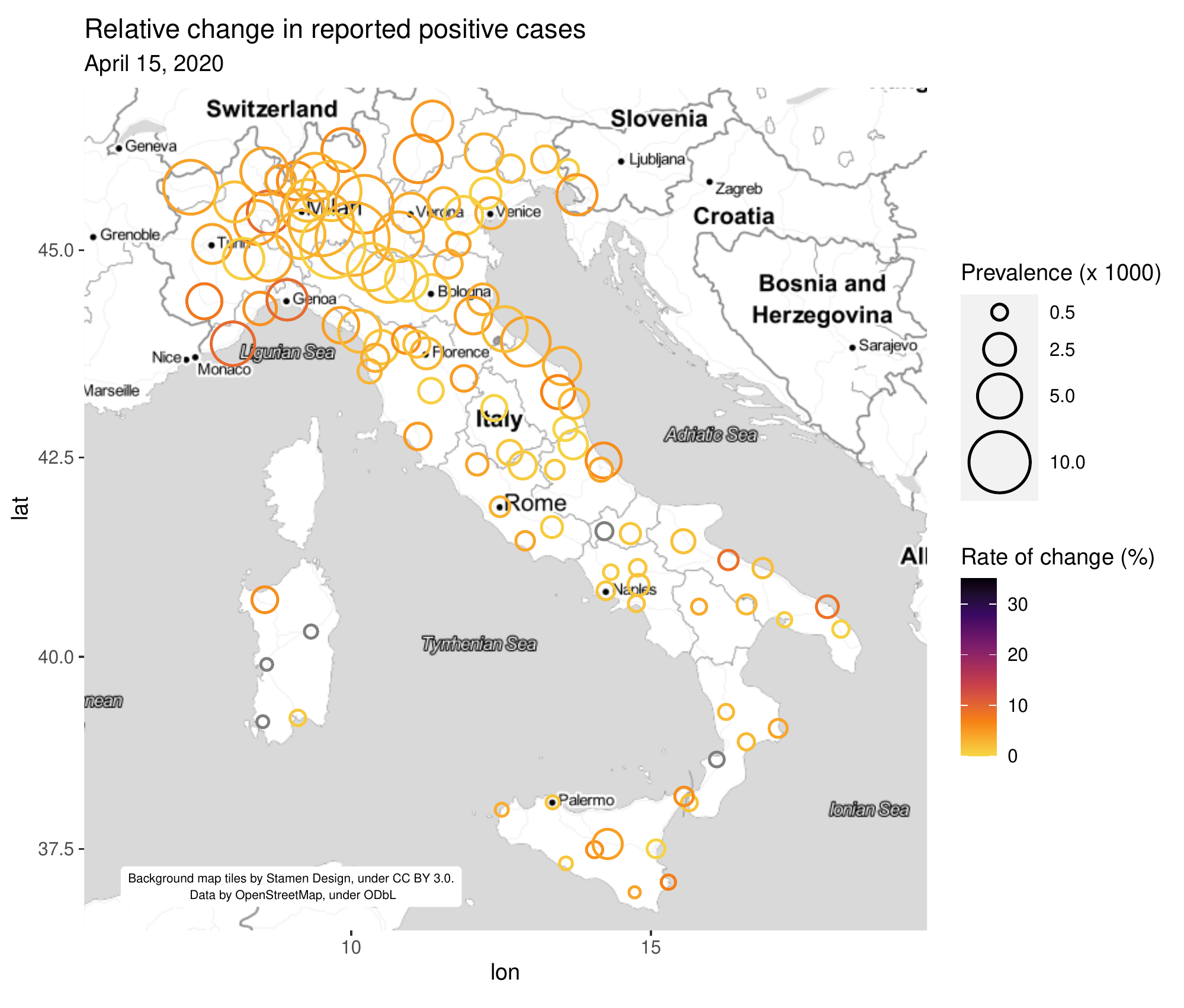}
\caption{{\small Map of Italy with circles indicating, for each province, the proportion of of reported infected SARS-CoV-2 cases (area of circle) and its rate of change (colors; see legend).}}
\label{fig:wholecountry}
\end{figure}

Figure \ref{fig:wholecountry} shows the whole country, with colored circles for the provinces as described in Section \ref{graphics}. This figure clearly shows that the current stage of the epidemic consists of a highly prevalent region in the Padana valley (with large circles) with relatively low rates of increase in the number of reported infected cases. Some of the peripheral ares have lower prevalence, but faster growth. In particular, the center-South and the islands are relatively silent, although a few provinces show a more rapid increase in their local epidemic.

\section{Conclusions}
It should be noted that the data that we have focused on here refers to the number of reported positives, which clearly underestimates the number of positives in the population. As such, the use of the terms prevalence and incidence would not be appropriate to use when analyzing these data. Note that one might consider the ratio of the reported positives to the number of performed tests instead. However, the process that leads to testing is not quite documented, and any changes in that strategy over time would make it difficult to interpret the estimated ratios. Note that such ratio should not be expected to be non decreasing, and that models for non-cumulative, binary data would seem more appropriate for such analyses.
\newline
Also, we are not using the date of onset of the infection (which in most cases is unknown), so that any inference will suffer from a delay relative to the trend of the infections in the population (see, e.g. \cite{wu:mcgo:2020}).
\newline
On the other hand, we believe that these analyses still have the potential for capturing trends that may be useful for the monitoring and management of the disease burden on the health care system. Our software also produces analyses for the other counts which are available, such as the number of Covid-positive deceased subjects and of Covid-positive patients in ICU and non-ICU departments at any given time. For the latter counts one clearly does not expect a monotonic trend in the observed counts, and as a consequence in our model selection process we did not prevent the algorithm to select as the best model a polynomial model that allows for a decrease in the rate, as we have done for the cumulative counts.

There are many possible alternative approaches to modeling these reported infections data. One option, which allows for forecasting from dynamic models, is described in \cite{chio:gaet:2002} and recently implemented at
\newline \mbox{\href{https://github.com/cgaetan/COVID-19}{https://github.com/cgaetan/COVID-19}}.
\newline
In addition, one may entertain models with spatial correlation across provinces, but the spread of the epidemic likely follows transportation routes rather than pure distance among provinces, so that it would not be trivial to implement such an extended, global model over a country. In addition, the starting times of the local epidemics in different regions appear to differ. Indeed, as shown in Figure 1, the choice of fitting separate regression models on the log scale estimated allows one to capture local epidemics that may (and do in our case) start at different times in different locations. The approach that we have followed here uses potentially different starting times for the different areas, while providing raw estimates of growth for the same (final) day.
\newline
The observed fit of the models seems very good. However, any models used in such ongoing monitoring will need to be assessed closely as the days go by, so that they may continue to provide a satisfactory fit for all provinces. In particular, the models for the cumulative counts should be closely monitored for their switching over time to higher degree polynomials and to the logistic curve formulation. Such counts can also be approached with other shapes of rescaled cumulative distribution functions, should the data suggest that. This would be particularly true if the data were to not satisfy the expected symmetry implicit in the logistic curve, on in other cumulative distribution functions. Alternatives could be the complementary log-log curve, or the generalized extreme value distribution. Curves containing some spline-based component may also be explored, but while these should be expected to provide excellent fit, their use in (limited) forecasting as we have performed here should be examined carefully to avoid its depending heavily only on the last few days of observed data 

Similarly to what was done for the local and global maxima, for the third-degree model one could estimate the time when the flex point was reached, or when it will be reached. From equation \ref{secder3} this requires solving a fourth-degree polynomial, which for simplicity one could solve numerically. We did not pursue this here.

Perhaps most importantly, the software runs in a few seconds. Repeating the analyses daily allows for the production of a new map for each day, and the comparison of such maps over time allows for the appreciation of the spatial dynamics of the epidemic across the provinces.

The \textsf{R} code and the file that contains the resident population of the Italian provinces are available from \href{https://sites.google.com/view/epilocal/home}{https://sites.google.com/view/epilocal/home}, or from the authors. A shiniyapp performing the analyses described in this article is available at https://ubasellini.shinyapps.io/testEPI/.
Comments and contributions are welcome.
\newline

\bibliography{covid}

\end{document}